\documentclass[useAMS,usenatbib]{mn2e}
\usepackage{txfonts}
\usepackage{natbib}
\usepackage{aas_macros}
\usepackage{graphicx}
\usepackage{subfigure}
\usepackage{mathrsfs}

\newcommand{\gsim}{\mathrel{\hbox{\rlap{\lower.55ex \hbox {$\sim$}}
                   \kern-.3em \raise.4ex \hbox{$>$}}}}
\newcommand{\lsim}{\mathrel{\hbox{\rlap{\lower.55ex \hbox {$\sim$}}
                   \kern-.3em \raise.4ex \hbox{$<$}}}}

\newcommand\sdensity{$\rm g \: cm^{-2}$}

\newcommand\rhill{$R_{\rm H}$}

\newcommand\rorbit{$r_{\rm p}$}

\newcommand\solarmass{$\rm M_{\odot}$}

\newcommand\earthmass{$\rm M_{\oplus}$}

\title[Planetary migration with steep temperature gradients]{Migration of protoplanets with surfaces through discs with steep temperature gradients}
\author[B.A. Ayliffe \& M.R. Bate]{Ben A. Ayliffe\thanks{E-mail:
ayliffe@astro.ex.ac.uk} and Matthew R. Bate\thanks{E-mail:
mbate@astro.ex.ac.uk}\\ School of Physics, University of Exeter, Stocker
Road, Exeter EX4 4QL \\ Centre for Stellar and Planetary Astrophysics \& School of Mathematical Sciences, Monash University, Melbourne Vic. 3800, Australia}

\date{\today}
\begin{document}
\maketitle

\begin{abstract}
We perform three-dimensional self-gravitating radiative transfer simulations of protoplanet migration in circumstellar discs to explore the impact upon migration of the radial temperature profiles in these discs. We model protoplanets with masses ranging between 10-100~\earthmass, in discs with surface density profiles of $\Sigma \propto r^{-1/2}$, and temperature profiles of the form $T \propto r^{-\beta}$, where $\beta$ ranges $0-2$. We find that steep ($\beta > 1$) temperature profiles lead to outward migration of low mass protoplanets in interstellar grain opacity discs, but in more optically thin discs the migration is always inwards. The trend in migration rates with changing $\beta$ obtained from our models shows good agreement with those obtained using recent analytic descriptions which include consideration of the co-orbital torques and their saturation. We find that switching between two models of the protoplanet, one in which accretion acts by evacuating gas and one in which gas piles up on a surface to form an atmosphere, leads to a small shift in the migration rates. If comparing these models in discs with conditions which lead to a marginally inward migration, the small shift can lead to outward migration. However, the direction and speed of migration is dominated by disc conditions rather than by the specific prescription used to model the flow near the protoplanet.
\end{abstract}

\begin{keywords}
planets and satellites: formation -- radiative transfer -- methods: numerical -- hydrodynamics -- planetary systems: formation
\end{keywords}

\section{Introduction}

Investigations into the interactions of circumstellar discs and embedded protoplanets were first conducted by \cite{GolTre1980}. They found  that the exchange of angular momentum between the two should lead to migration of the protoplanet (latterly known as Type I migration), but their work did not suggest in which direction this migration would be. The analysis of \cite{War1986} for a non-self gravitating, two-dimensional disc, showed that if the disc had a negative temperature gradient (i.e. temperature reduces with increasing heliocentric radius) then an embedded protoplanet should migrate inwards. When Hot Jupiters were discovered \citep{MayQue1995} and the difficulties with their in situ formation became apparent, this inward migration seemed to be an ideal mechanism by which to reconcile existing formation scenarios with these planets' small final orbital radii. However, the timescales of growth and migration did not at first compare favourably. Migration timescales were so short that protoplanets should plummet into their stars before they were able to grow to any considerable mass. Only by reaching masses sufficient to open disc gaps can protoplanets move from fast Type I migration to slower Type II migration, which proceeds at the disc's viscous evolution timescale. Once slowed, the planets only have to survive until the disc gas dissipates, thought to occur at a stellar age of less than 6 Myrs \citep*{HaiLadLad2001}, at which point migration due to planet-gas interactions necessarily ceases.

Since \cite{War1986} further analytical descriptions  (i.e. \citealt*{War1997, TanTakWar2002}), and numerical modelling  (\citealt{KorPol1993}; \citealt{NelPapMasKle2000}; \citealt{Mas2002}; \citealt*{DAnHenKle2002}; \citealt{BatLubOgiMil2003}; \citealt*{DAnKleHen2003}; \citealt*{AliMorBen2004}; \citealt*{DAnBatLub2005};  \citealt{KlaKle2006}; \citealt{DAnLub2008}; \citealt{LiLubLiLin2009}) of planet migration have continued to find fast inward rates in the Type I regime ($<$ 100 \earthmass). However, these works have generally considered locally-isothermal conditions. A large number of works, both analytical and numerical, have now been devoted to exploring the impact of more complex thermodynamics upon planet migration. The first of such works was published by \cite{MorTan2003}, who performed local calculations focussing on a low-mass planet's interaction with its parent disc. They found that radiative cooling led to a non-axisymmetric mass distribution about the planet. This resulted in an additional torque beyond the commonly considered differential Lindblad torques, and altered the migration rate relative to a purely isothermal calculation. \cite{MenGoo2004} found that a planet's migration rate was sensitive to the temperature and opacity structure of the disc through which it travelled, and under certain conditions the migration rate could be very much slower than that achieved in an isothermal model. This was closely followed by \cite{JanSas2005} who also identified a slow down in Type I migration rates upon the introduction of radiative transfer.

It was \cite{PaaMel2006} who first found cases of outward migration due to a coorbital torque in their three-dimensional radiative transfer models. They found that this coorbital torque could be prevented from saturating if the radiation diffusion timescale was shorter than the libration timescale of gas in the horseshoe region. This enabled temperature asymmetries to be maintained beyond a single libration period. They also showed that reducing the opacity (further shortening the radiation diffusion timescale) of their disc could lead to results in agreement with previous isothermal calculations. This was further described by \cite{BarMas2008}, who make clear that the radiation diffusion timescale must be greater than the period of a single horseshoe orbit to avoid returning to an isothermal like migration, and that outward migration can be related to the disc's radial entropy gradient. Many numerical models have since also found evidence of outward migration in the Type I regime (\citealt{PaaMel2008}; \citealt{PaaPap2008}; \citealt{KleCri2008}; \citealt*{KleBitKla2009}; \citealt{AylBat2010}; \citealt{YamIna2011}). Periods of outward migration can help to increase the overall migration timescale of a forming planet embedded in a disc, as is required by synthesis models to explain the population of exoplanets that has been observed \citep*{IdaLin2008, MorAliBen2009}.

More recently there has been an extension of analytical descriptions of the non-linear coorbital torque, called the horseshoe drag, to describe it in both its unsaturated and saturated states (\citealt{PaaBarCriKle2010}; \citealt*{PaaBarKle2011}; \citealt{MasCas2010}). These works give expressions for the total torque acting on a planet due to both Lindblad torques and the coorbital component. As such they can describe planet migration for a large range of scenarios, taking into account the disc's density and temperature profiles, as well as its thermal diffusivity. Outward migration is fastest in discs with steep radial temperature profiles, becoming more marginal at typical gradients such as $T \propto r^{-1}$.

This paper is intended to explore protoplanet migration in a series of discs with radial temperature gradients from $r^{0}$ to $r^{-2}$. We conduct three dimensional global disc models, using smoothed particle hydrodynamics (SPH), that include self-gravity and which use a planetary surface to allow modelling of gas flow to well within the Hill sphere and the self-consistent formation of an atmosphere. We conduct a few isothermal models to investigate whether or not the temperature profile alone has an impact on the migration rate, but otherwise our calculations all include radiative transfer using a flux-limited diffusion approximation. In Section~\ref{sec:setup}, we describe our computational method, in Section \ref{sec:calc} we explain how we obtained our results which are then presented in Section~\ref{sec:results}. Section~\ref{sec:discussion} discusses these results, whilst a summary and our conclusions are given in Section~\ref{sec:summary}.

\section{Computational Method}
\label{sec:setup}

The calculations described herein have been performed using a three-dimensional SPH code. This SPH code has its origins in a version first developed by \citeauthor{Ben1990} (\citeyear{Ben1990}; \citealt{BenCamPreBow1990}) but it has undergone substantial modification in subsequent years. Energy and entropy are conserved to timestepping accuracy by use of the variable smoothing length formalism of \cite{SprHer2002} and \cite{Mon2002} with our specific implementation being described in \cite{PriBat2007}. Gravitational forces are calculated and neighbouring particles are found using a binary tree. Radiative transfer is modelled in the two temperature (gas, $T_{\rm g}$, and radiation, $T_{\rm r}$) flux-limited diffusion approximation using the method developed by \citet*{WhiBatMon2005} and \citet{WhiBat2006}.  Integration of the SPH equations is achieved using a second-order Runge-Kutta-Fehlberg integrator with particles having individual timesteps \citep{Bat1995}. The code has been parallelised by M. Bate using OpenMP.

\subsection{Equation of state and radiative transfer}
\label{sec:RT}

We present a few calculations performed using a locally-isothermal equation of state, as well as many more calculations which include radiative transfer. In locally-isothermal models the temperature of the gas in the disc remains as a function of radius throughout the calculations. For the radiation hydrodynamical calculations we use the ideal gas equation of state, $p=\rho T_{g} R_{g}/\mu$ where $R_{g}$ is the gas constant, $\rho$ is the density, $T_{g}$ is the gas temperature, and $\mu$ is the mean molecular mass. The equation of state takes into account the translational, rotational, and vibrational degrees of freedom of molecular hydrogen (assuming a 3:1 mix of ortho- and para-hydrogen; see \citealt{BolHarDurMic2007}). It also includes the dissociation of molecular hydrogen, and the ionisations of hydrogen and helium.  The hydrogen and helium mass fractions are $X=0.70$ and $Y=0.28$, respectively, whilst the contribution of metals to the equation of state is neglected.  More details on the implementation of the equation of state can be found in \cite{WhiBat2006}.

The radiative transfer in these calculations is performed using the flux-limited diffusion approximation, as implemented by \cite{WhiBatMon2005} and \cite{WhiBat2006}, in which work and artificial viscosity (including both bulk and shear components) increase the thermal energy of the gas. Work done on the radiation field increases the radiative energy which can be transported via flux-limited diffusion. The energy transfer between the gas and radiation fields is dependent upon their relative temperatures, the gas density, and the opacity, $\kappa$. Energy is lost from the system by the radiation field into the vacuum surrounding the disc; this is performed numerically through the use of a radiation boundary. This boundary is positioned at a height from the disc midplane at which the radiation path through the gas from this height outwards has an optical depth of $\tau_{\rm op} \approx 1$; i.e. the radiation can be expected to reach the vacuum. As a result, two layers of particles, one above and one below the disc, are made to comprise the radiation boundary. These particles are compelled to follow the initial temperature profile of the disc, causing them to function as energy sinks which impose a minimum temperature to which the disc can cool \citep[see][for a fuller description]{AylBat2010}.

The use of steep radial temperature profiles in some of our models in this work leads to very high temperatures towards the inner boundary of these circumstellar discs. This material forms an unrealistically hot annulus at the inner boundary, from which radiation diffuses out through the disc, changing its structure out to the environs of the embedded protoplanet. To remedy this situation, the disc out to 2 au is compelled to maintain its initial temperature profile throughout the simulation, which quashes any diffusion in this region. This rule was applied in all the calculations for consistency, including the shallower temperature profile calculations where it has no discernible effect.

The opacities used here are those of  \cite{PolMcKChr1985} and  \cite{Ale1975}  (the IVa King model), with the former providing the grain opacities and the latter the gas opacities at temperatures beyond the grain sublimation point. In some calculations the grain opacity is reduced by a factor of 100 to emulate possible modification of the population due to agglomeration processes \citep[see][for further details]{AylBat2009}.

\subsection{Disc models}

We model a protoplanetary disc with radial bounds of 0.1 - 3 \rorbit \ (0.52 - 15.6 au), where \rorbit \ is the initial orbital radius of our embedded protoplanets, taking a value of 5.2 au. The disc is represented by 2 million SPH particles, a number found to deliver satisfactory resolution \citep[see][]{AylBat2010}. This leads to smoothing lengths at the disc midplane at \rorbit \ of $H/4$ (away from the planet), where $H$ is the disc scaleheight and equals 0.05 at \rorbit. Of course, considerably better resolution is obtained directly surrounding the protoplanet where the densities become much higher. The surface density of the disc goes as $\Sigma \propto r^{-1/2}$, where the value at \rorbit \ is $\Sigma_{\rm p} =$ 75 \sdensity \ to allow comparison with previous works (e.g. \citealt*{LubSeiArt1999}; \citealt{BatLubOgiMil2003}). At the centre of the disc is a fixed potential representing a 1~\solarmass \ star. We implement an inner boundary for the disc that prevents gas flow onto the star, which due to the lack of a self-limiting mechanism would otherwise artificially rapidly drain material from the disc. The outer edge of the disc is bordered by ghost particles which represent the disc beyond 15.6 au, preventing the disc from shear spreading into a vacuum \citep[for more details see][]{AylBat2010}. The initial discs were evolved in the absence of a planet until any transience resulting from settling had dissipated, which required just over 4 orbits of the disc's outer edge. Self-gravity is included in all the calculations described in this paper. The impact of including self-gravity was discussed in \cite{AylBat2010}, where for these relatively low mass discs it was found to make only a marginal difference to migration rates, slightly slowing inward migration. However, self-gravity is essential for building a self-consistent atmosphere within the Hill radius.

\begin{figure}
\centering
\includegraphics[width=1.0 \columnwidth]{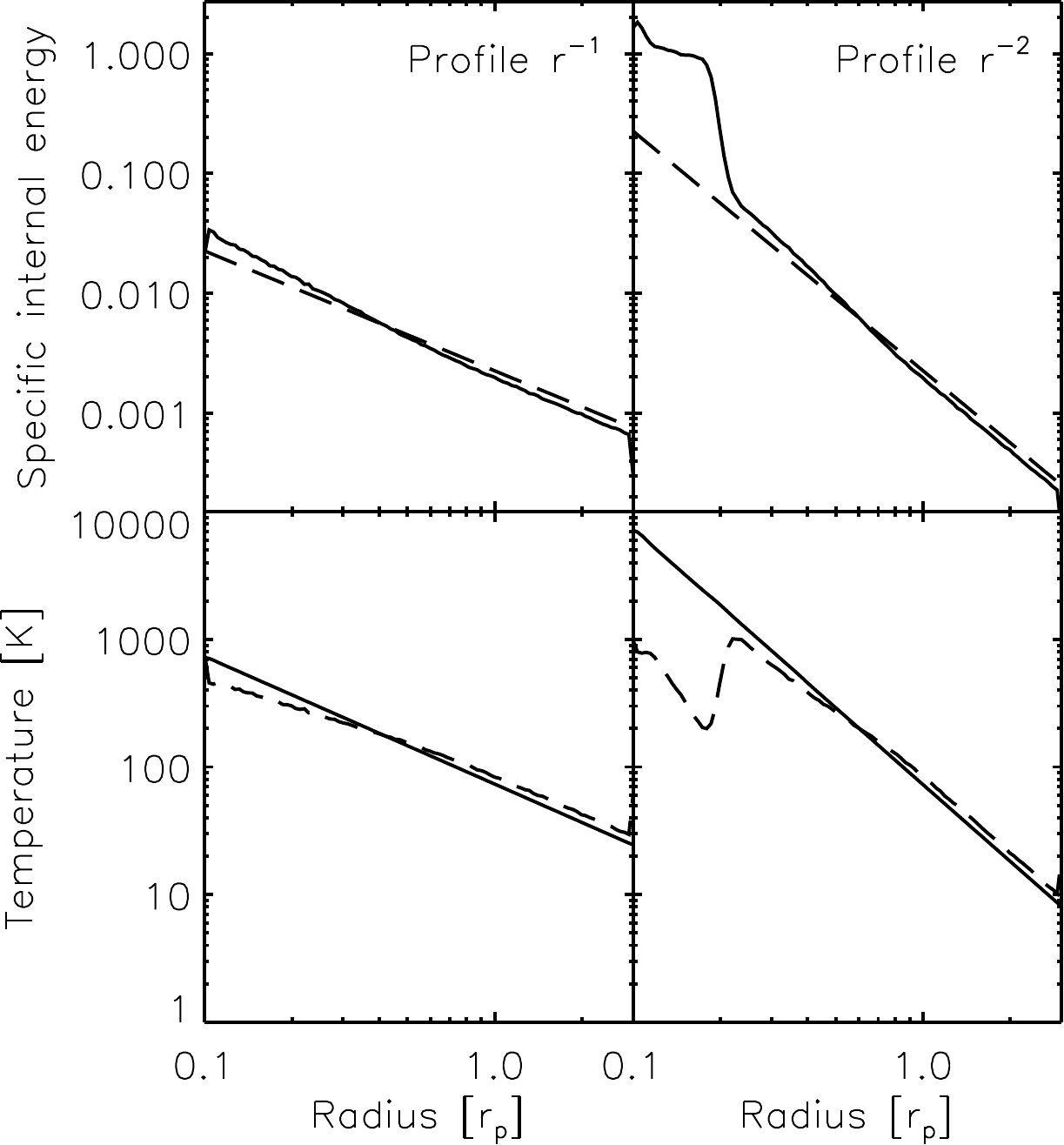}
\caption{Internal energy ($u$, top panels) and temperature ($T$, bottom panels) profiles along the midplane of unperturbed discs. The dashed lines show the profiles that are given when it is $u$ that is established with the given radial dependence. Conversely, the solid lines show the profiles that are produced when it is $T$ which is compelled to follow the stated profile. For the shallow profile given in the left panels, the temperature profile that results from setting $u \propto r^{-1}$ is similar to that obtained by forcing $T \propto r^{-1}$. However, where a steeper profile is used, as in the right panels, the temperatures reached are sufficient to bring about substantial changes in the specific heat capacity of the gas (i.e. due to dissociation of molecular hydrogen, excitation of rotational and vibrational modes of molecular hydrogen) such that the proportionality of energy and temperature breaks down; we use heat capacities calculated using \protect \cite{BolHarDurMic2007}. As such in this work it is the temperature profile that is set directly ($T \propto r^{-\beta}$), for which the corresponding energy is then found.}
\label{fig:uvstemp}
\end{figure}

We perform calculations with a number of radial temperature profiles described by $T \propto r^{-\beta}$ where $\beta = $ 0, 0.5, 1, 1.5, and 2, and $T_{\rm p} \approx$ 73 K. Due to the use of steeper temperature profiles in this work, which results in higher temperatures at smaller heliocentric radii, the method used to set the disc temperatures is different to that used in \cite{AylBat2010}. In this previous work the initial temperature distribution was set assuming that the heat capacity at constant volume was independent of temperature, such that $T \propto u$, where $u$ is the gas's specific internal energy. This is a good approximation so long as the rotational and vibrational modes of $\rm H_{2}$ are not excited (i.e. $\gamma = 5/3$), which is true for $T \lesssim 100$K. It was then the distribution of $u$ that was actually prescribed to follow an $r^{-1}$ profile, such that $\beta \approx 1$; this case is shown by the dashed lines in the left panels of Fig.~\ref{fig:uvstemp}. The resulting temperature profile is similar to the desired $r^{-1}$ profile as can be seen by comparing the solid line in the lower left panel, for which $T$ is set explicitly, with the dashed line that results from setting $u$. The higher temperatures in some of the disc models considered in this work fall beyond the range in which this assumption can readily be made, as shown by the right hand panels of Fig.~\ref{fig:uvstemp}. As such, for these models we set the initial disc temperatures and boundary temperatures to follow the desired profile (the solid lines in lower panels of Fig.~\ref{fig:uvstemp}) and then the appropriate internal energy for the gas is found (solid lines in top panels of Fig.~\ref{fig:uvstemp}) using tabulated heat capacities based upon \cite{BolHarDurMic2007} which are used in our radiative transfer implementation. We note that at the planet's orbital radius the disc's temperature is such that $\gamma \approx 5/3$. Therefore we adopt this value for $\gamma$ when making comparisons with analytic models.

The viscosity within the disc is introduced in a parameterised form, developed for SPH by \cite{MonGin1983}, which conserves linear and angular momentum, and was modified to deal with high Mach number shocks by \cite{Mon1992}. It is a function of two parameters, the $\alpha_{\textsc{sph}}$-term establishes a viscosity to damp subsonic velocity oscillations that may be produced in the wake of a shock front, whilst the $\beta_{\textsc{sph}}$-term prevents particle interpenetration in supersonic shocks. This viscosity is only applied when the gas is under compression, and should ideally only act near shock fronts. We use the switch developed by \cite{MorMon1997} to try to reduce the action of artificial viscosity where the cause is not a shock. This attributes an  $\alpha_{\textsc{sph}}$-value to each particle, which is modified based on the local pressure gradient, such that it increases to a maximum of 1 at a shock, and falls away rapidly to 0.1 as the particle moves away; this significantly reduces unwanted viscous dissipation in the disc.

\section{Calculations}
\label{sec:calc}

\subsection{Planet model and migration measurements}

Into the protoplanetary discs we embed a protoplanetary core, or partially formed protoplanet. In all the calculations discussed here the protoplanet is represented by a point mass which is free to move, and in all but a few cases (discussed below) the point mass is enwrapped by a surface of radius 0.03 times the Hill radius (\rhill) \citep{AylBat2009, AylBat2010}. Gas is able to pile up on this surface to build an atmosphere, and so represent gas accretion in a reasonably realistic manner. As in our previous work, a smooth start to the calculations is achieved by embedding a protoplanet of radius $R_{\rm p} =0.01 r_{\rm p}$ which then shrinks exponentially to the desired radius during the first orbit of the protoplanet.

The point mass is free to move through the disc, enabling us to measure its rate of migration. This measurement is made by using a linear fit to the last 25 orbits of orbital radius evolution, in what is typically a 50 orbit total evolution. The first 25 orbits are given up to ensure that any transient disruption to the disc caused by the sudden introduction of a planet have subsided, and to allow gas in the horseshoe region to reach a quasi-equilibrium state. For a 10~\earthmass \ planet the libration time is $\approx 50$ orbits, for a 333~\earthmass \ planet it is $\approx 8$ orbits, whilst for the a 33~\earthmass \ planet it is 29 orbits. As such, our lowest mass protoplanet mass models have completed a single libration period at the end of the calculations. In \cite{AylBat2010} we presented our results using migration timescales, $\tau$, which are calculated as the time a planet of fixed mass would take to migrate from its starting orbital radius (5.2 AU in all of the models discussed) in to the central star; $\tau = r_{\rm p}/\dot{r}$.  This has no real meaning if the protoplanet is migrating outwards, so in this paper we present results using simply the migration rate, $\dot{r}$, in units of $r_{\rm p}$ per orbital period at $r_{\rm p}$ (which for $r_{\rm p} = 5.2$~au is 11.86 years).

We further refine the presentation of our results by quantifying the uncertainties that arise due to our method of measuring the migration rates. To achieve this, in addition to fitting over the entirety of the last 25 orbits, we also divide this interval into 3 smaller overlapping time domains to which individual linear fits are made. For a migration trend that is well represented by a linear fit, these 3 migration rates should all be similar to both one another and to the rate measured over the entire 25 orbits. For less linear trends the measured rates give an indication of the possible variation that could be obtained by measuring at different points in the orbital evolution. We use the maximum and minimum migration rates obtained from these shorter time domain fits to provide error bars for each migration rate that is presented in this work.

In \cite{AylBat2010} we noted a difference between the migration rates measured when modelling a protoplanet with the surface described above and an accreting point mass. To explore this further we conduct a small number of calculations in which the protoplanet is represented as a point mass which accretes gas that comes within $r_{\rm acc}$, where $r_{\rm acc} = 0.03 R_{\rm H}$ in each case. The particles representing this accreted gas are removed from the calculation and their properties are used to modify those of the point mass, ensuring conservation of mass, linear momentum, and angular momentum \citep*[see][]{BatBonPri1995}.

\subsection{Opacity to diffusivity}
\label{sec:diffusivity}

\cite{MasCas2010} developed an analytic expression to calculate the torques acting on protoplanets embedded in discs with temperature profiles described by the relation $T = T_0 (r/r_p)^{-\beta}$, such as the discs we have modelled in this work. They also include terms that describe the impact on the torques of the disc's thermal diffusivity, which in our radiative transfer calculations is effectively varied through changes in the opacity. In order that we might compare our numerical models with \citeauthor{MasCas2010}'s analytic expression, we calculate the thermal diffusivity in our disc at \rorbit \ in the absence of a protoplanet. In the flux-limited diffusion approximation of radiative transfer used in this work the flux in optically thick regions is given by

\begin{equation}
\mbox{\boldmath $F$} = - \frac{c}{3 \chi \rho} \nabla E,
\label{eq:fluxeq}
\end{equation}

\noindent where $c$ is the speed of light, $\chi$ is the Rosseland mean opacity, $\rho$ is the gas density, and $E = a T_r^{4}$ in which $a$ is the radiation density constant, $a = 4 \sigma_{\textsc {sb}}/c$. This allows us to restate equation \ref{eq:fluxeq} as

\begin{equation}
\mbox{\boldmath $F$} = - \frac{16 \sigma_{\textsc {sb}} T_r^3}{3 \chi \rho} \nabla T_r.
\label{eq:fluxtemp}
\end{equation}

\noindent \citeauthor{MasCas2010} give an expression for the heat flux in terms of a form of thermal diffusivity, $k$, as

\begin{equation}
\mbox{\boldmath $F$} = -k \nabla T.
\label{eq:massetflux}
\end{equation}

\noindent Comparing equations \ref{eq:fluxtemp} and \ref{eq:massetflux} we obtain an expression for $k$, which is related to a true thermal diffusivity, $\kappa$, by

\begin{equation}
\kappa = \frac{\gamma -1}{\gamma} \frac{k \mu}{R_g \rho},
\label{eq:diffusivityconvert}
\end{equation}

\noindent which is similar to equation 33 in \cite{MasCas2010}, where $\gamma$ is the adiabatic index. The difference between this equation and that of \citeauthor{MasCas2010} lies in their use of $\Sigma$ in the denominator of equation~\ref{eq:diffusivityconvert} as they are describing a two-dimensional disc, whilst for three-dimensions we use $\rho$. Making use of the approximation that $\rho = \Sigma/H$ gives our final equation

\begin{equation}
\kappa = \frac{\gamma -1}{\gamma} \frac{16 \sigma_{\textsc {sb}} T_r^3 \mu H}{3 \chi \rho R \Sigma},
\label{eq:diffusivity}
\end{equation}

\noindent where $H$ has a value of 0.05 at \rorbit \ regardless of the chosen disc temperature profile. With this equation, taking $\gamma = 5/3$ (which is appropriate at \rorbit), and values from our unperturbed initial disc we are able to calculate the thermal diffusivities in models of differing opacity, $\chi$. For the 100\% interstellar grain opacity disc this gives $\kappa = 6.57 \times 10^{15} {\rm cm^{2}s^{-1}}$, and the 1\% opacity disc has a value of $\kappa = 6.57 \times 10^{17} {\rm cm^{2}s^{-1}}$. These diffusivities correspond with characteristic disc cooling timescales of tens of orbital periods and less than an orbital period, respectively, where this timescale is linearly dependent upon the chosen diffusivity. These timescales indicate that in the higher diffusivity (lower opacity) calculations the imposed boundary will dominate the temperature structure of the disc, but this is not the case in the lower diffusivity (higher opacity) models.

It may also be of note that in these three-dimensional models, the dominant direction for radiative diffusion within the horseshoe region is vertically. For the high opacity calculations, vertical diffusion is found to be more than three times as efficient as radial diffusion.

\section{Results}
\label{sec:results}

\subsection{Is the surface treatment important?}

\begin{figure}
\centering
\includegraphics[width=0.95 \columnwidth]{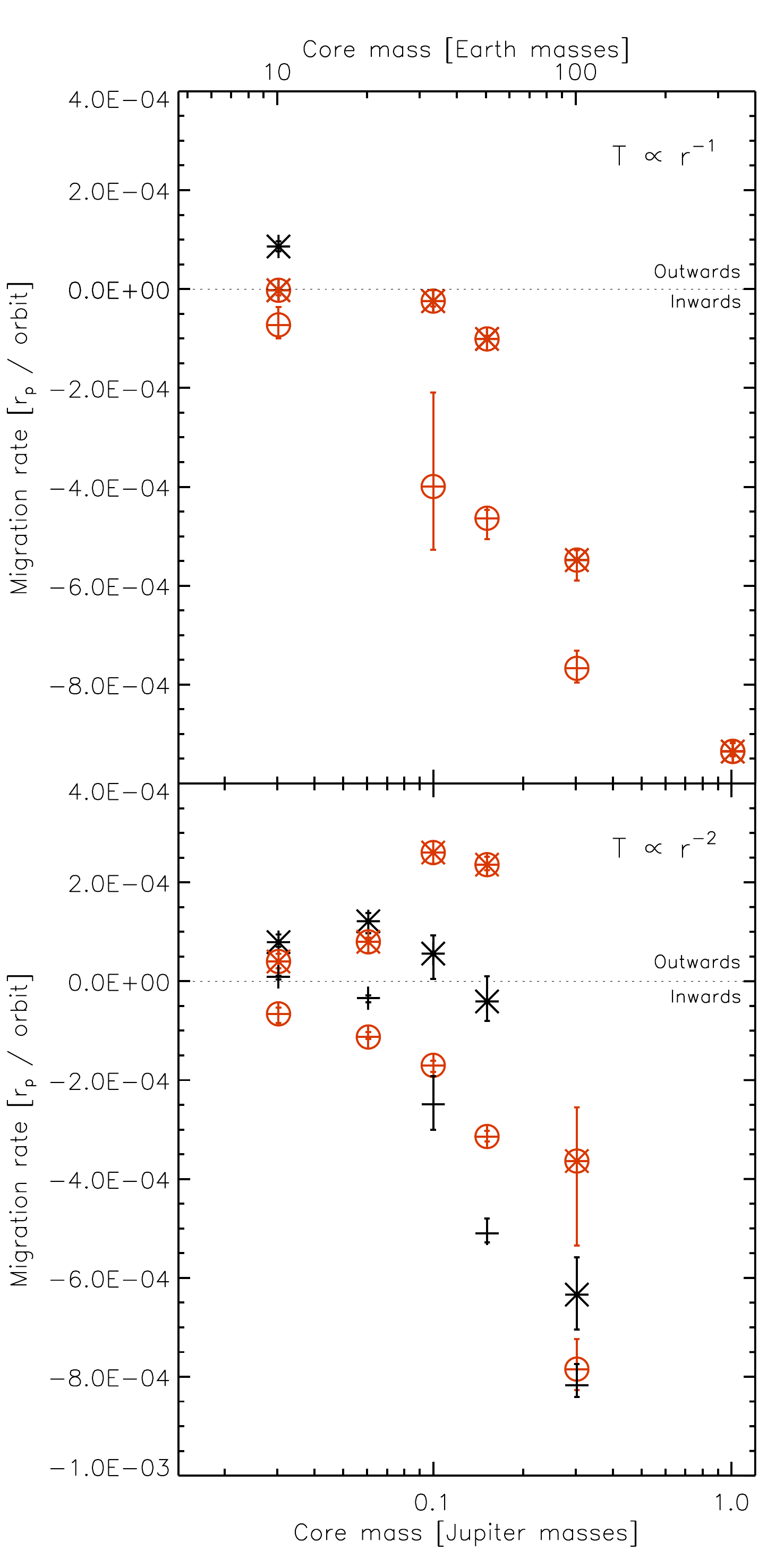}
\caption{The upper panel shows migration rates, which were presented as migration timescales in \protect \cite{AylBat2010}, for protoplanets modelled using surfaces (symbols encircled, red) embedded in discs with $\beta \approx 1$. These discs were of either 100\% (asterisks) or 1\% (plus symbols) interstellar grain opacity. For a 333~\earthmass \ protoplanet changing the opacity has such a small impact upon migration that the two rates appear indiscernibly overlaid above. The lower panel shows migration rates for protoplanets embedded in a circumstellar disc for which $\beta = 2$, some modelled with evacuating sinks (un-encircled symbols, black) and others using sinks with surfaces (once again, encircled symbols, red). All subsequent figures include only protoplanets modelled using sinks with surfaces.}
\label{fig:nosurface}
\end{figure}

First we address a comment made in \cite{AylBat2010} in which we suggested that the modelling of migrating protoplanets using a surface treatment rather than an accreting point mass might significantly alter their migration rates, even changing the direction. This was inspired by a single result for a 10~\earthmass \ protoplanet in a disc with a temperature profile exponent of $\beta \approx 1$, in which the migration was seen to change from outward to inward when a crude evacuating sink particle was replaced with a protoplanet with a surface in a high opacity disc (shown in the top panel of Fig.~\ref{fig:nosurface}). To explore this further we have conducted a series of calculations spanning $10 - 100$~\earthmass \ in a disc with a steeper temperature profile, $\beta = 2$. It was envisaged that such a steep profile would lead to outward migration for a low mass protoplanet modelled with either of our treatments by increasing the significance of the coorbital torques, making their effects more apparent. The results of these calculations are shown in the lower panel of Fig.~\ref{fig:nosurface}. It can be seen that using 1\% interstellar grain opacities (plus symbols), even with this steep temperature profile, the protoplanet migration is predominantly inwards, with the 10~\earthmass \ case being the only exception in exhibiting an almost stalled outwards migration. However, at 100\% interstellar grain opacities those protoplanets with masses $\lesssim 33$~\earthmass \ are all migrating outwards, regardless of the chosen protoplanet treatment. Comparing the migration of these low mass protoplanets with those seen in the top panel of Fig.~\ref{fig:nosurface} illustrates the increased importance of the coorbital torques in the disc with a steeper temperature gradient.

In our new calculations (bottom panel) the 10~\earthmass \ and 20~\earthmass \ protoplanet migration rates obtained with an evacuating sink (un-encircled, black) are more rapidly outwards or less rapidly inwards than their counterparts modelled using sinks with surfaces (encircled, red); the reason for this is unclear, though it is in agreement with the sole case shown in the top panel for our previous calculations. However, at masses $\gtrsim 33$~\earthmass \ this pattern reverses regardless of the disc opacity. \cite{BatLubOgiMil2003} showed that it was for such masses that a protoplanet might begin to form a non-negligible gap in the disc. The evacuation of a gap reduces the mass in the coorbital region and as a result reduces the magnitude of coorbital torques. In \cite{AylBat2010} we found that between the two treatments of the protoplanet used here, it was the evacuating sink particles that developed clearer gaps as a result of drawing in material artificially fast relative to protoplanets modelled with surfaces. This suggests that protoplanets of $\gtrsim 33$~\earthmass \ modelled with evacuating sinks migrate faster inwards (slower outwards) than there equivalents modelled with surfaces because they reduce the coorbital torques, which act outwards, to a greater extent. For a 100~\earthmass \ protoplanet the coorbital region is sufficiently depleted with both treatments to lead to inward migration in a high opacity disc.

As to our findings in \cite{AylBat2010} that changing the protoplanet treatment can reverse the direction of migration for a low mass protoplanet, we must clarify that this appears to be a result of the marginal nature of the migration of a 10~\earthmass \ protoplanet in a high opacity disc with $\beta \approx 1$. The change in the forces acting when comparing a protoplanet modelled using an evacuating sink and one modelled using a surface, happened to be sufficient in this case to change the direction of migration, though more typically it is likely to give a different rate with the same direction (as seen in Fig.~\ref{fig:nosurface}). The essential point is that as a protoplanet's migration rate is found to be at least somewhat dependent upon the method by which its accretion is modelled, it is best that the adopted treatment be as realistic as possible. We therefore remain convinced that using a surface to allow gas-pile-up as a model of accretion is the best method for these purposes.

\begin{figure*}
\centering
\includegraphics[width=1.0 \columnwidth]{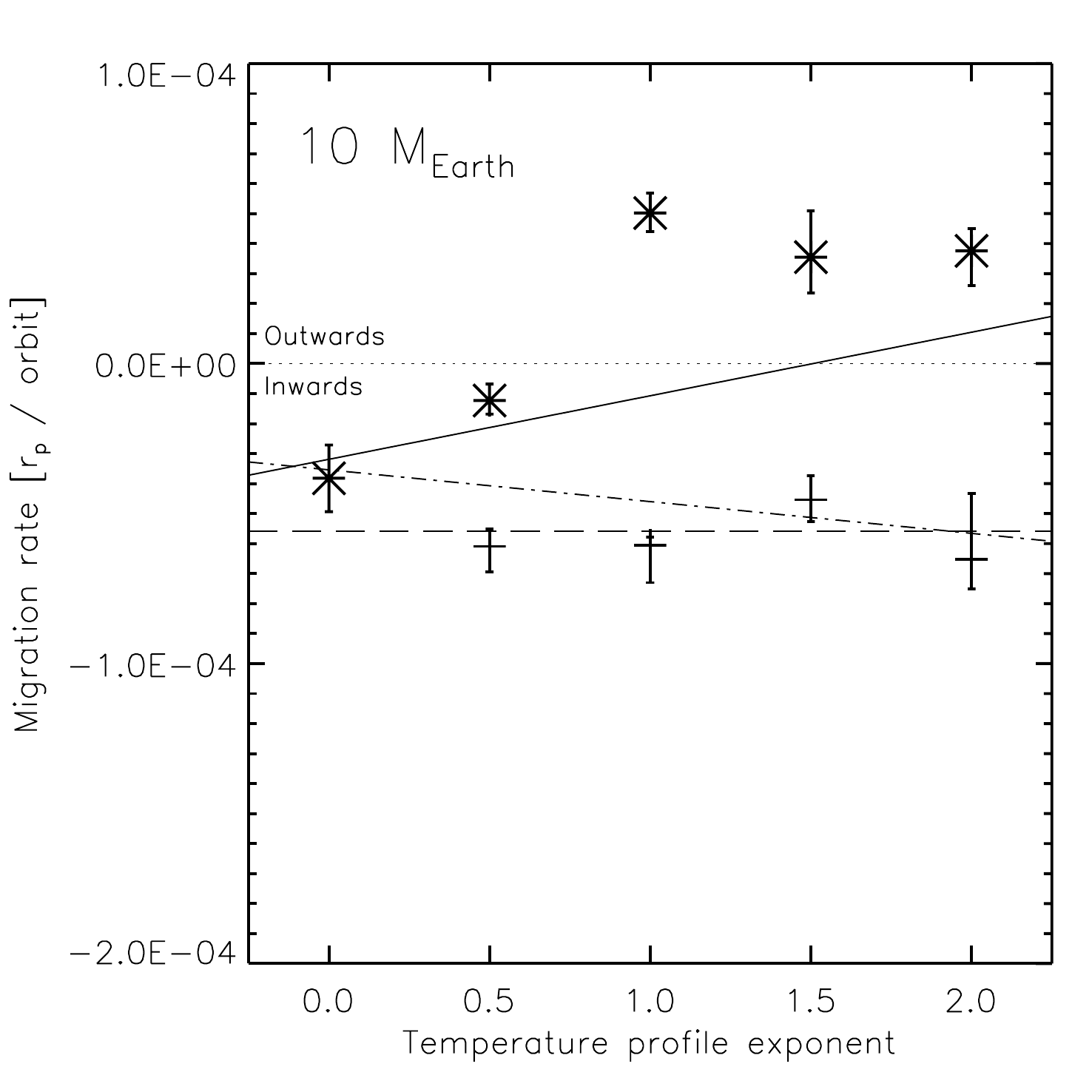}
\includegraphics[width=1.0 \columnwidth]{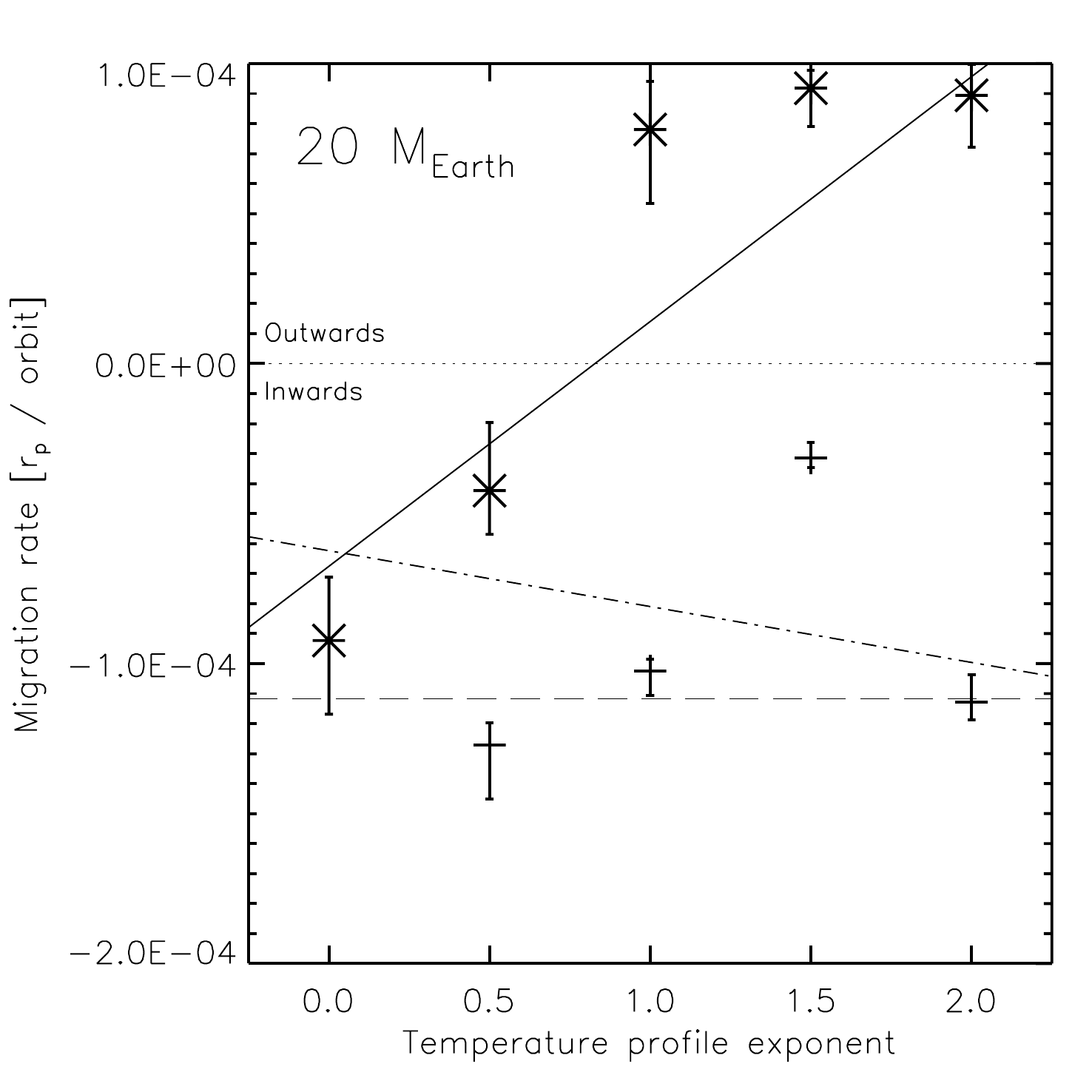}
\includegraphics[width=1.0 \columnwidth]{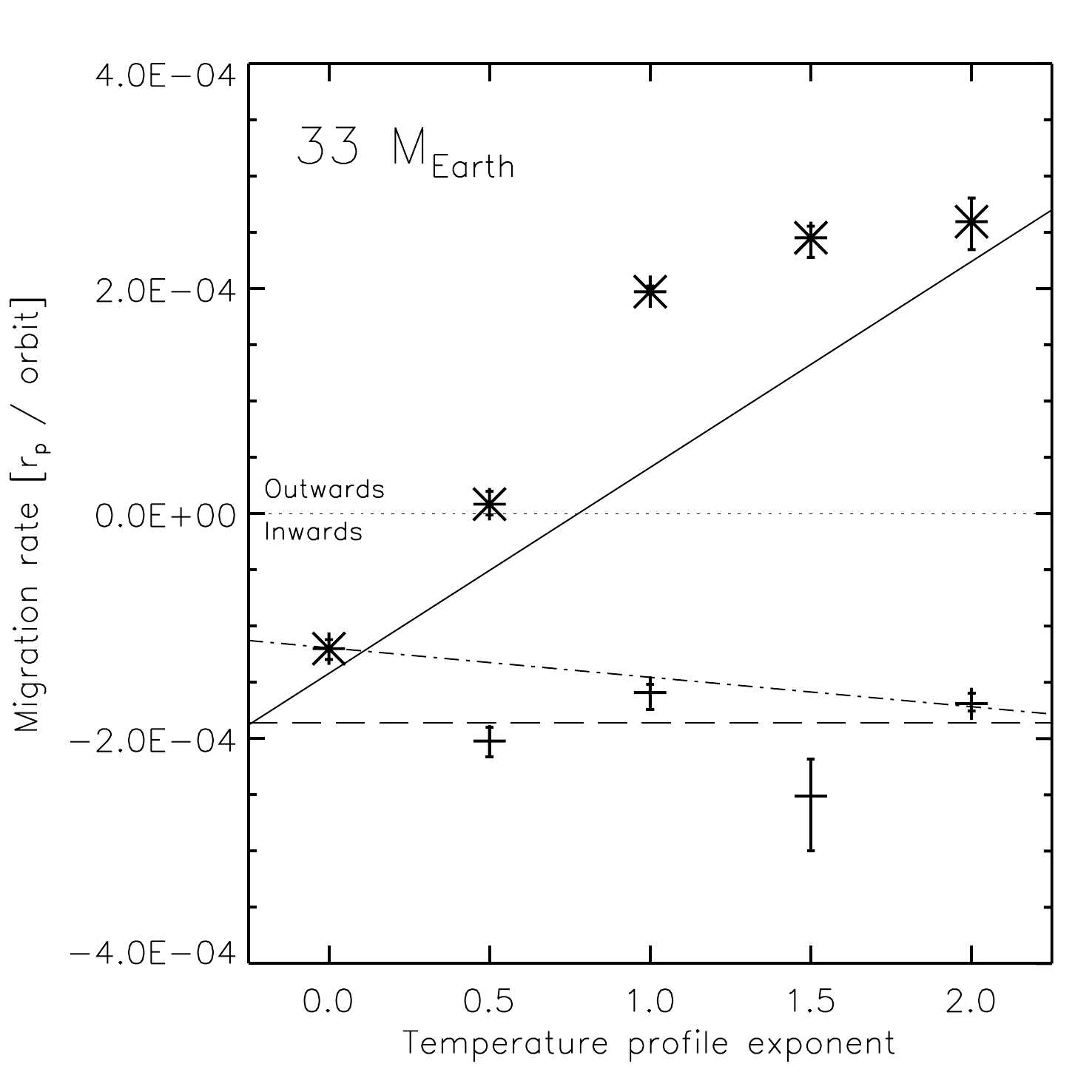}
\includegraphics[width=1.0 \columnwidth]{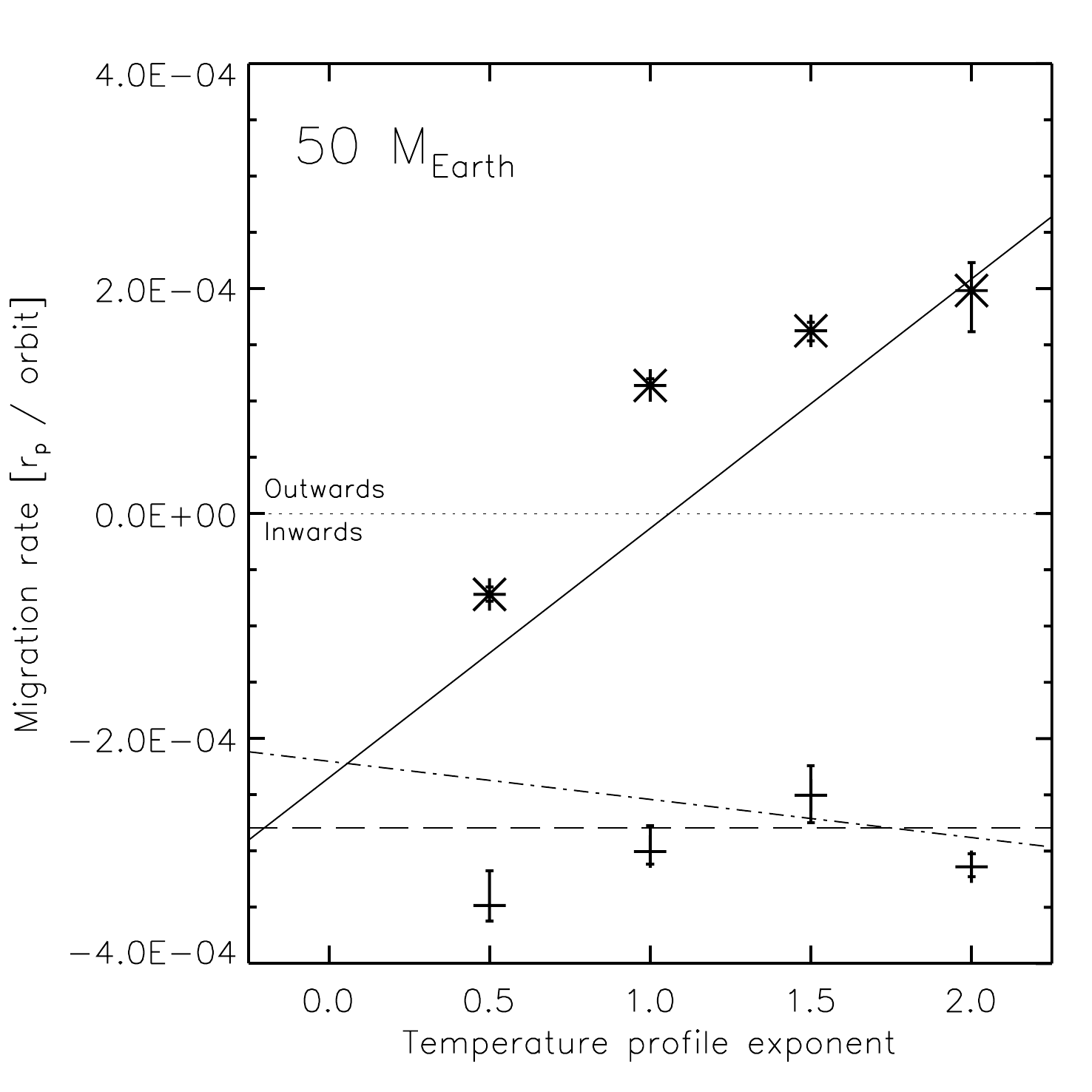}
\caption{Migration rates of four different mass protoplanets; across and down the panels 10, 20, 33, and 50 \earthmass \ (as marked). Each protoplanet was modelled in a range of circumstellar discs with different temperature profiles, $T \propto r^{-\beta}$, where the temperature profile exponent $\beta$ is marked on the x-axis. Models in which interstellar grain opacities are used (asterisks) lead to slower or reversed migration when compared with the 1\% interstellar grain opacity models (plus symbols). \protect \cite{MasCas2010} give an expression for the total torque acting upon a protoplanet which includes the dependency upon the disc's temperature profile and its thermal diffusivity, as well as the planet's mass. The resulting migration rates are plotted in each panel, one for a diffusivity akin to our 100\% opacity calculation (solid line), and one for the 1\% opacity case (dot-dash line). Migration rates obtained from \protect \cite{TanTakWar2002} are independent of the temperature profile, and so appear in each panel as a horizontal dashed line, the rate of which is determined by the disc surface density and the protoplanet mass.}
\label{fig:profiles}
\end{figure*}

\subsection{Impacts of the temperature profile}

Our primary goal in this work was to explore the impact of the temperature profile in a circumstellar disc upon the migration of protoplanets therein. In particular we wished to test the predictions of recent analytic models, which by their nature are unable to include details of the gas flow near the protoplanet; Fig.~\ref{fig:profiles} contains the results of our simulations to this end. Shown are the migration rates for protoplanets with masses 10, 22, 33, and 50~\earthmass, embedded in discs with both high and low opacities, and a range of values of $\beta$. A horizontal dashed line in each panel marks the migration rate obtained using \cite{TanTakWar2002}'s analytic form, which depends on the disc surface density and protoplanet mass, but is independent of the value of $\beta$. It can be seen that the low opacity models (1\% interstellar grain opacity, marked with plus symbols) display some scatter, but generally centre around this analytic description. All of the low opacity calculations lead to inward migration, as would be expected for locally-isothermal calculations.

Also marked in Fig.~\ref{fig:profiles} are analytic migration rates based on recent work by \cite{MasCas2010}. In their description there is a dependence on both $\beta$ and on the thermal diffusivity of the disc. As such we plot lines to correspond with both opacities used in calculating our results, where the diffusivities used are those calculated in Section \ref{sec:diffusivity}; note that high opacities correspond to low thermal diffusivities and vice versa. In the case of the high opacity calculations (100\% interstellar grain opacity, marked with asterisks) it can be seen that as the steepness of the temperature profile is increased, the inward migration rates slow towards zero before the direction changes to outwards and the rates begin to increase. In each panel of Fig.~\ref{fig:profiles}, excepting the 10~\earthmass \ case for which the uncertainties are greatest, it is apparent that the rate of this change is in good agreement with the gradient given by the low thermal diffusivity (high opacity) result of \citeauthor{MasCas2010}'s analytic description (solid line). The analytic form also gives reasonable agreement (given the coarse spacing of $\beta$ in our results) as to the temperature profile at which a given mass protoplanet in a high opacity disc will transition from inwards to outwards migration. The formula of \citeauthor{MasCas2010} is also dependent upon the disc's viscosity, which for plotting here we have used a \cite{ShaSun1973} $\alpha$-value of 0.004, which is equivalent to the magnitude of the SPH viscosity in the vicinity of the protoplanet in our models. Note also that this analytic description is valid only up to masses for which the half-width of the horseshoe region, $x_s$, is less than the disc thickness at \rorbit. Adopting the adiabatic approximation of \cite{BarMas2008} for $x_s$ gives a maximum mass of 40~\earthmass, meaning that the 50~\earthmass \ cases shown in Fig.~\ref{fig:profiles} are slightly beyond the range of validity.

\begin{figure}
\centering
\includegraphics[width=1.0 \columnwidth]{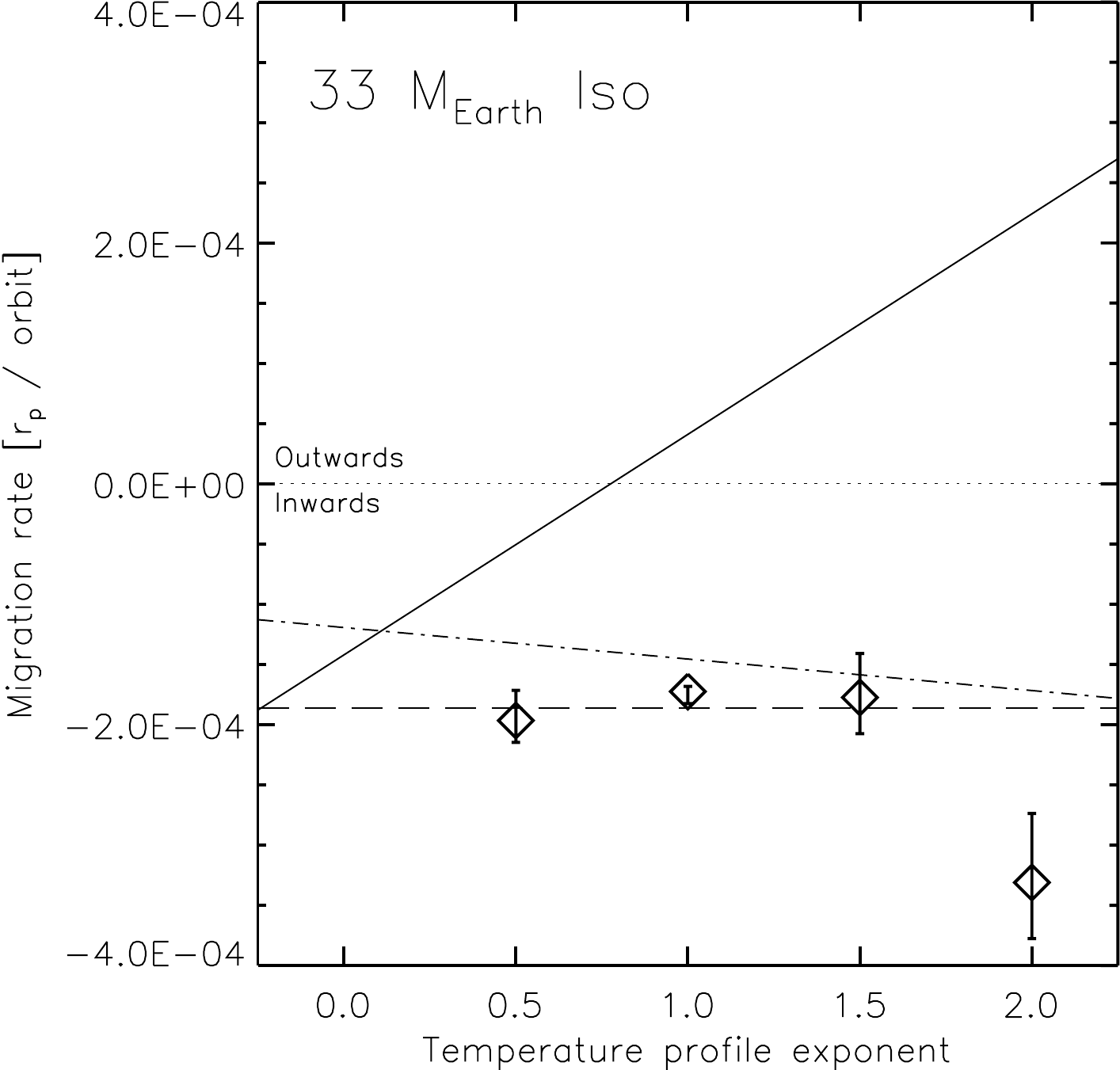}
\caption{Migration rates for a 33~\earthmass \ protoplanet in true locally-isothermal discs with varying radial temperature profiles. There is a small range of migration rates, varying by less than a factor of two from the minimum to the maximum rate. No significant trend is evident.}
\label{fig:isothermal}
\end{figure}

Using \citeauthor{MasCas2010}'s analytic form with values that correspond to our low opacity models indicates a reversal of the migration trend with increasing $\beta$, resulting in migration accelerating inwards for steeper profiles. However, the scatter in migration rates measured from our low opacity calculations is large, and prohibits us from concluding anything other that that the magnitudes show reasonable agreement. True locally-isothermal models were also conducted to establish whether they revealed a trend with the disc temperature profile. The results of these models are shown in Fig.~\ref{fig:isothermal}, where once again there appears to be no definite trend in the change of migration rates with increasing $\beta$, and the variation in the rates is small, varying by less than a factor of 2 from the minimum to maximum rate. As such we are unable to say whether or not the migration trend at low opacities given by \cite{MasCas2010} is real.

\begin{figure}
\centering
\includegraphics[width=1.0 \columnwidth]{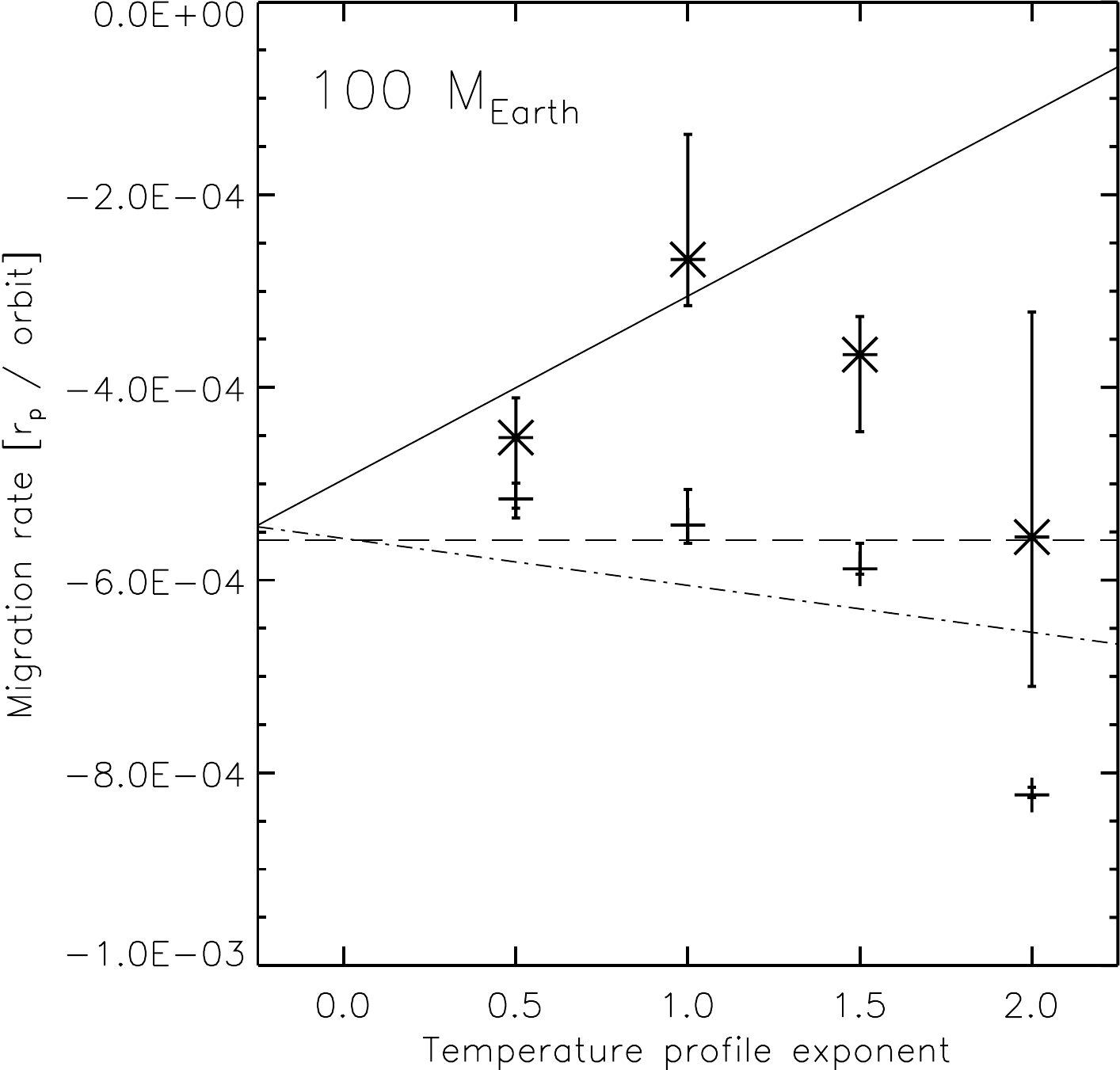}
\caption{Migration rates of a 100~\earthmass \ protoplanet in a range of circumstellar discs with different temperature profiles, $T \propto r^{-\beta}$, where the temperature profile exponent $\beta$ is marked on the x-axis. Models in which interstellar grain opacities are used (asterisks) generally lead to very slightly slower inward migration than their equivalent 1\% interstellar grain opacity (plus symbols) models. However, the trend with $\beta$, which was quite pronounced for the high opacity models of lower mass protoplanets, is more subdued in this case. It also fails to follow the trend predicted using \protect \cite{MasCas2010} for a high opacity disc (solid line), though at 100~\earthmass \ we are applying the analytic model well beyond its range of validity; the analytic trend for a low opacity disc is marked with a dot-dashed line. The change in disc opacity leads to a smaller change in the protoplanet migration rate for this relatively high mass protoplanet when compared with the lower mass protoplanets shown in Fig.~\ref{fig:profiles}. Rates from \protect \cite{TanTakWar2002} are shown as a horizontal dashed line.}
\label{fig:100m}
\end{figure}

Considering the immutable properties of our disc model, we find that only protoplanets $< 100$~\earthmass \ may migrate outwards for high values of $\beta$. This can be seen in Fig.~\ref{fig:100m} where migration rates for a 100~\earthmass \ protoplanet are plotted, and it is clear that no combination of opacity and $\beta$ is able to cause outward migration. Included in the figure are the same analytic descriptions used in Fig.~\ref{fig:profiles}, but here the \cite{MasCas2010} line is well beyond the mass regime for which it is valid. For such a high mass protoplanet, the evacuation of the coorbital region should effectively end the significant influence of the coorbital torques. As such these effects are overestimated by the analytic model, which is consistent with the fact that our high opacity results all lie well below the solid line. Noticeably the analytic model, even with the overestimation of the coorbital torques which should favour outward migration, does not predict such migration for this high mass protoplanet (for $\beta \lesssim 2.5$) , suggesting it is not possible in discs such as those modelled here.

\subsection{Impact of absolute temperature}

\begin{figure}
\centering
\includegraphics[width=1.0 \columnwidth]{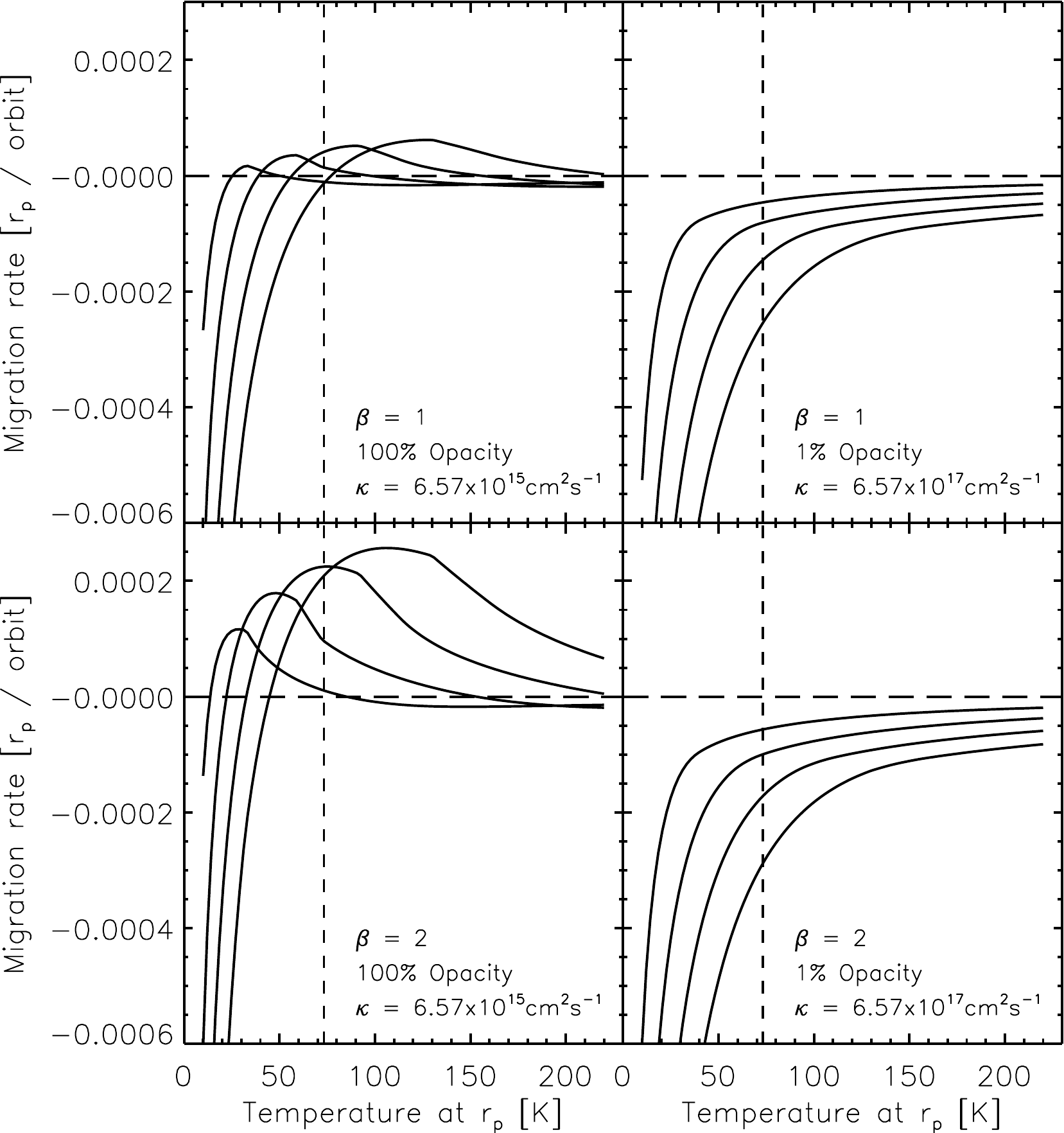}
\caption{Migration rates of protoplanets in discs of differing absolute temperature, calculated using the analytic description of \protect \cite{MasCas2010}. Rates calculated for conditions equivalent to our 100\% opacity models are shown on the left hand side, and 1\% opacity on the right. The temperature profiles are $T \propto r^{-1}$ and $r^{-2}$ in the top and bottom rows respectively. The four lines in each panel from left to right correspond to protoplanets with masses 10, 20, 33, and 50~\earthmass, respectively. The vertical dashed line marks the temperature at $r_{\rm p}$ used in our numerical models.}
\label{fig:abstemp}
\end{figure}

In this paper, we have studied the impact of the disc's temperature profile on migration rates. In all these calculations, the absolute disc temperature at the planet's orbital radius is the same. As discussed above we find reasonable agreement with the analytic description of \citeauthor{MasCas2010}. However, using \citeauthor{MasCas2010}'s analytic description it is possible to examine the migration rate's dependence upon the absolute temperature at $r_{\rm p}$; this is presented in Fig.~\ref{fig:abstemp}. The migration rates, calculated assuming a disc with a diffusivity of $\kappa = 6.57 \times 10^{15} {\rm cm^{2}s^{-1}}$ (equivalent to our 100\% opacity cases), are shown in the left hand panels. For each of the four protoplanet masses plotted ($m_{\rm p} = $ 10, 20, 33, 50~\earthmass) there is a temperature (scaleheight) at which the outward migration rate achieves a maximum, with this temperature scaling as $m_{\rm p}^{0.85}$ (within the mass range for which the model is valid, $\lesssim 40$~\earthmass). The right hand panels of Fig.~\ref{fig:abstemp} show the same cases for a higher diffusivity of $\kappa = 6.57 \times 10^{17} {\rm cm^{2}s^{-1}}$ (equivalent to our 1\% opacity cases), for which conditions there is no outward migration, regardless of the chosen absolute temperature of the disc, and there is no turnover in the rate of migration, though these rates still vary significantly with temperature.

These analytic results imply that as well as the temperature profile, the chosen initial disc temperature is an important factor in determining the migration behaviour of the various low mass planets. Temperatures in different discs around various types of star may be expected to exhibit temperatures throughout the range shown here of 10-230K at 5.2~au, suggesting further modelling should be carried out for a range of disc conditions to explore the utility of analytic descriptions like \cite{MasCas2010}, which if valid, can be applied more generally than the results of numerical modelling.

\section{Discussion}
\label{sec:discussion}

\begin{figure}
\centering
\includegraphics[width=1.0 \columnwidth]{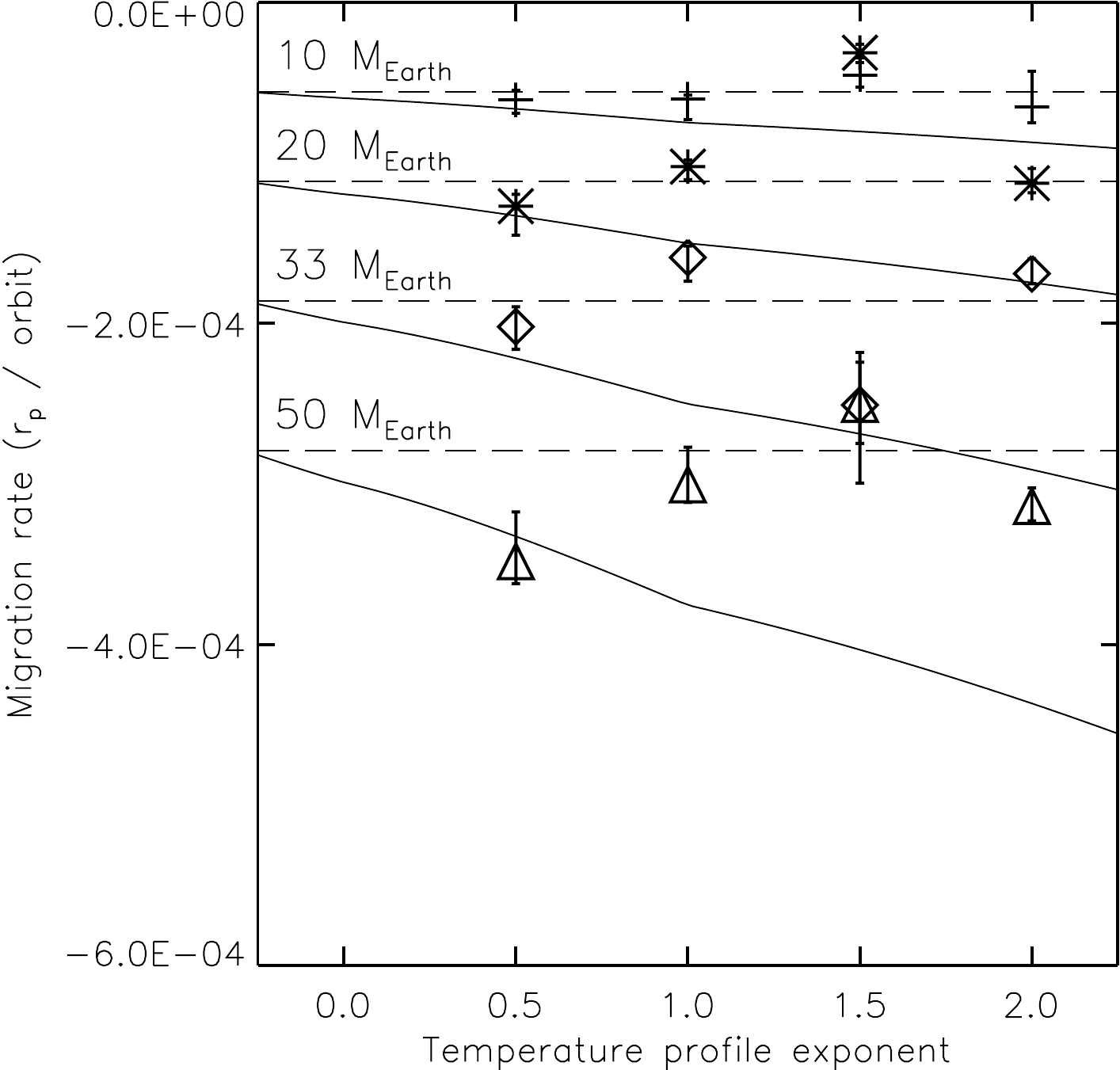}
\caption{Migration rates of four different mass protoplanets through low opacity discs of differing temperature profiles. The masses are 10~\earthmass \ (plus symbols), 20~\earthmass \ (asterisks), 33~\earthmass \ (diamonds), and 50~\earthmass \ (triangles). Also included are analytic predictions. Those of \protect \cite{TanTakWar2002} without a $\beta$ dependence, for a three-dimensional isothermal disc, appear as horizontal dashed lines. The solid lines show the rates predicted using the $\beta$ dependent expression devised by \protect \cite{DAnLub2010} for a three-dimensional locally-isothermal disc. The masses that these lines correspond to are marked on the plot above the corresponding pairs.}
\label{fig:lubow}
\end{figure}

First we address the low opacity calculations that we have performed. The resulting migration rates from these calculations appear scattered, showing no evident trend with changes in the disc temperature profile. This can be seen in Fig.~\ref{fig:lubow} which gathers all the low opacity results shown in Fig.~\ref{fig:profiles} into a single panel, as well as including the analytic rates for an isothermal disc calculated using \cite{TanTakWar2002} (dashed lines) and rates based upon the recent work of \cite{DAnLub2010} (solid lines) for a locally-isothermal disc. At these low opacities we expect the migration results from our models to be similar to those predicted with isothermal analytic expressions as we found in \cite{AylBat2010}. Indeed, despite the scatter, these migration results do reveal a trend with increasing mass that is matched by \cite{TanTakWar2002}'s values. However, \citeauthor{DAnLub2010} have found that in the locally-isothermal (rather than globally isothermal) regime protoplanet migration rates depend upon the parent disc's temperature profile, with steeper negative temperature profiles (higher values of $\beta$) leading to more rapid inward migration. This trend is not seen in our low opacity calculation results within which, as mentioned previously, it is not possible to establish a trend that is distinguishable from the numerical uncertainties; this is also the case in the locally-isothermal calculations shown in Fig.~\ref{fig:isothermal}. However, the variation in migration rates obtained from \cite{DAnLub2010} across the range of $\beta$ considered in this work is relatively small, less than a factor of 2 for each of the masses shown in Fig.~\ref{fig:lubow}. Our rather large uncertainties prevent us from definitively ruling out the model.

\begin{figure}
\centering
\includegraphics[width=1.0 \columnwidth]{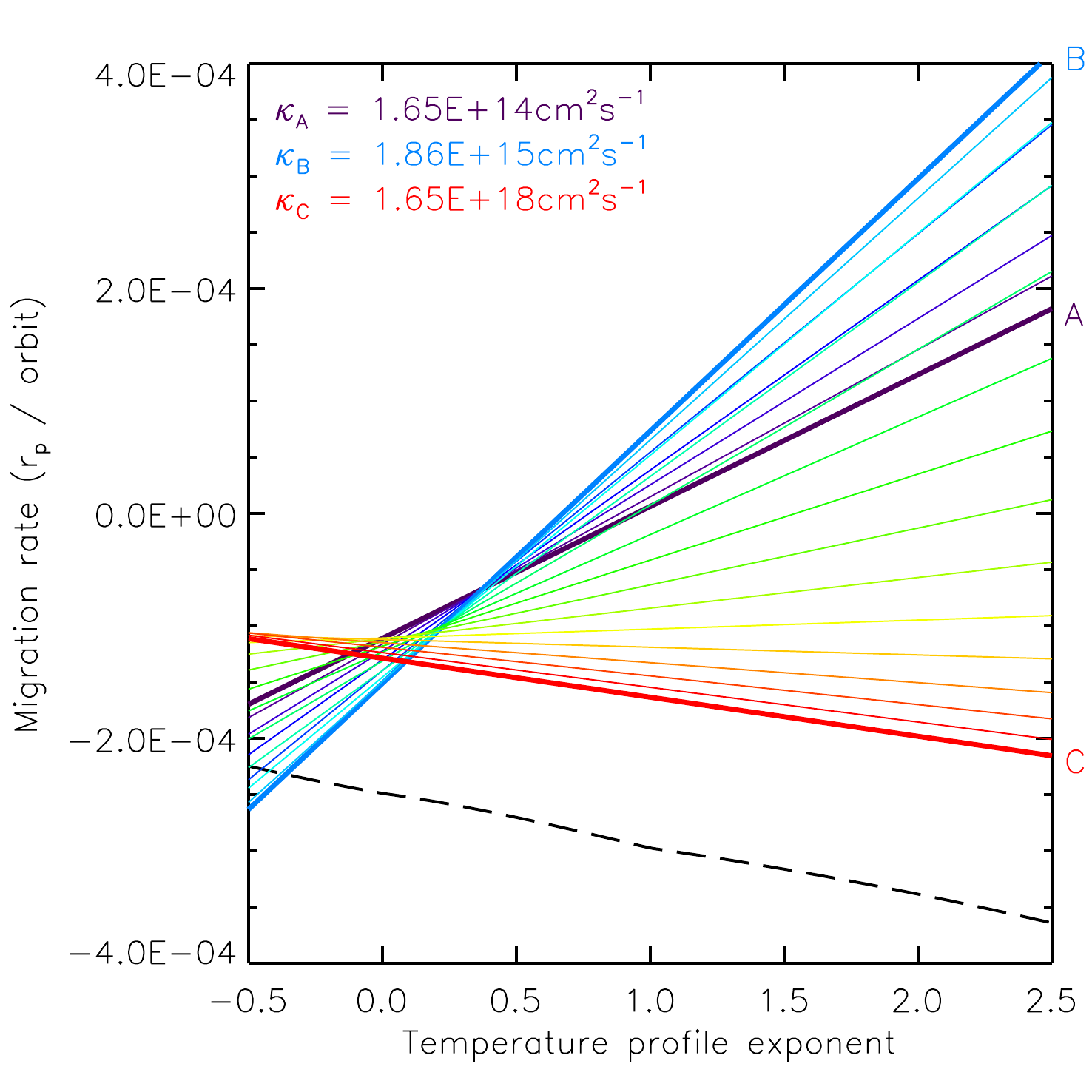}
\caption{Migration rates based on the total torques calculated using analytic expressions for a 33~\earthmass \ protoplanet in a disc with properties equivalent to our models. The dashed line indicates the expected locally-isothermal result for such a protoplanet in a three-dimensional disc, obtained using \protect \cite{DAnLub2010}. The solid lines are for rates obtained using \protect \cite{MasCas2010}, where the colours progressing through the rainbow correspond to differing diffusivities, with the highest shown in red (labelled C). Three labelled thicker lines identify the trend with diffusivity, with the corresponding values given in the key. The locally-isothermal and highest diffusivity cases share the same gradient, with magnitudes differing by $\approx 5/3$. }
\label{fig:lubmas}
\end{figure}

Returning to \cite{DAnLub2010}'s analytic form for an isothermal disc, it is interesting to note that the acceleration of inward migration with increasing $\beta$ is in qualitative agreement with the description of \cite{MasCas2010} discs with high thermal diffusivity. As a brief aside we compare these two analytic models with which we draw comparisons. Fig.~\ref{fig:lubmas} illustrates the migration rates calculated using \citeauthor{DAnLub2010}'s isothermal formula for a 33~\earthmass \ protoplanet (dashed line), with the rates obtained using \citeauthor{MasCas2010}'s formula for a whole range of thermal diffusivities also shown. It can be seen that at the highest diffusivity (thick red line labelled C) the gradient of the change of migration rate with increasing $\beta$ for the two descriptions is extremely similar, and they are displaced in magnitude by a factor of~$\approx 5/3$, equivalent to $\gamma$, as was found in \cite{PaaPap2008}. This similar behaviour is to be expected, with a high thermal diffusivity leading to a disc that resembles a locally-isothermal model, and responds to different temperature profiles in the same manner. As seen earlier when considering our low opacity calculations, the manner in which the migration rate of a protoplanet changes with the discs thermal diffusivity is dependent upon the temperature profile of the disc. For discs with $\beta \gtrsim 0.5$, as the thermal diffusivity is increased from its lowest value (thick purple line labelled A) the migration rate of a protoplanet can be expected to initially increase in the outward direction, or slow down if the migration is inwards. Beyond a certain diffusivity, shown by the thick blue line labelled B in Fig.~\ref{fig:lubmas}, the trend is reversed with increasing diffusivity slowing outward migration, and accelerating inward migration. This behaviour is a result of the edge term of the Horsehoe drag in \citeauthor{MasCas2010}'s model. In future it would be desirable to explore a broader opacity space with greater refinement using our numerical models to determine whether we can capture the effects of the edge terms of the Horseshoe drag.

\begin{figure*}
\centering
\includegraphics[width=2.0 \columnwidth]{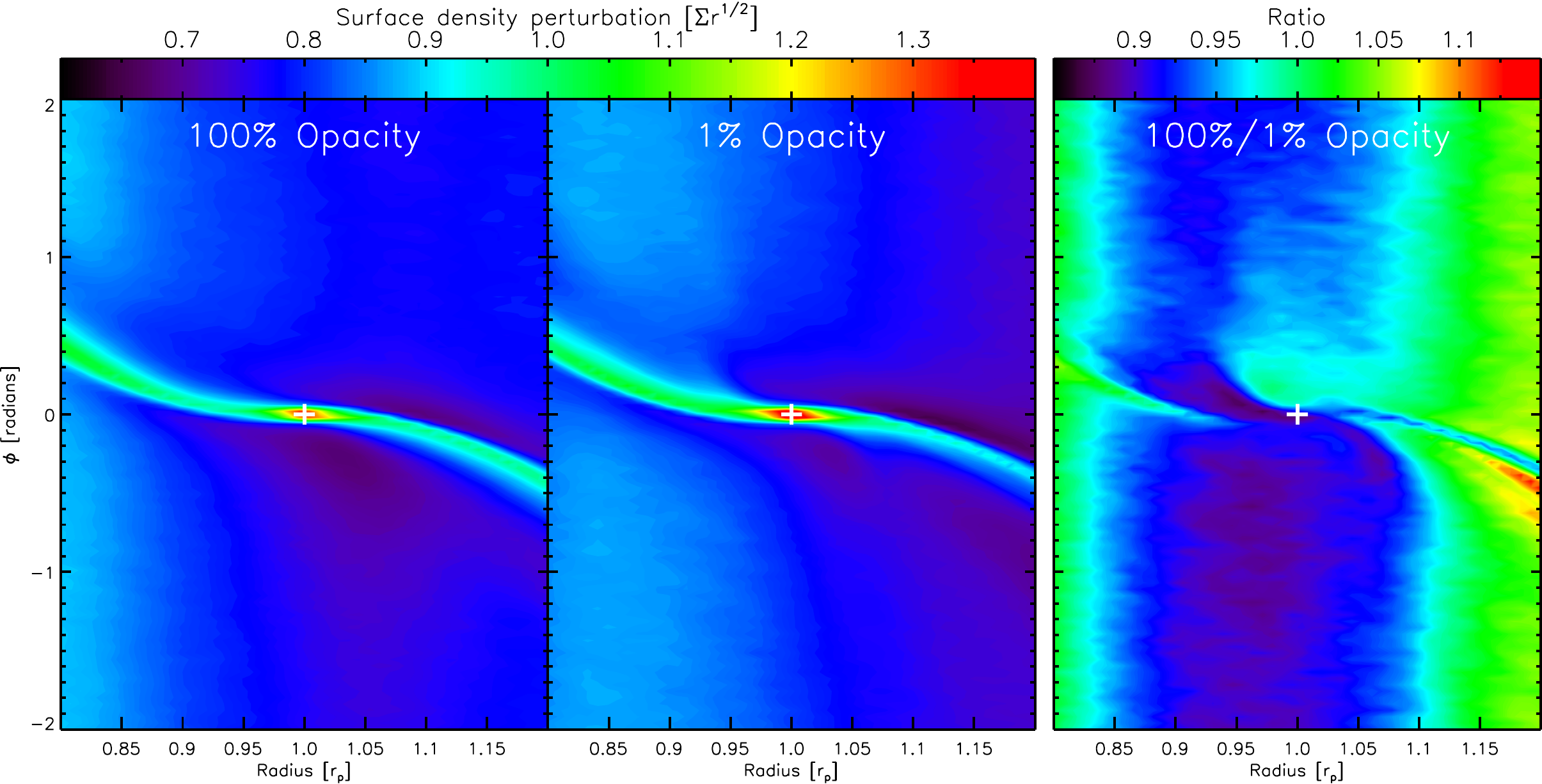}
\caption{The two left panels show the surface density normalised by the density gradient of the initial unperturbed disc. Respectively the panels show the 100\% and 1\% opacity models with $\beta = 2$ and a protoplanet mass of 33~\earthmass. The right hand panel shows the ratio of the two left panels, revealing how the structures differ between the optically thick and thin models. This makes it clear that relative to the low opacity (inward migrating) model the high opacity (outward migrating) model has a greater contrast between leading (positive $\phi$) and trailing (negative $\phi$) densities, thus explaining the differing directions of migration.}
\label{fig:cylindrical}
\end{figure*}

In our high opacity (low thermal diffusivity) models we find that increasing $\beta$ from 0 leads first to slower inward migration and then accelerating outward migration for protoplanets of $\lesssim 50$~\earthmass. This is in agreement with the analytic work discussed above. Similar outward migration has been found in the numerical models of \cite{PaaPap2008}, \cite{BarMas2008}, \cite{KleBitKla2009}, and \cite{PaaPap2009}, and has been associated with a density enhancement ahead of the planet and a deficit behind it in the horsehoe region. In each of these models the feature has been clearly identified in two-dimensional modelling, but such a feature is likely to be much less apparent in three-dimensional models where gas is not constrained to move in a single plane. The lefthand panels of Fig.~\ref{fig:cylindrical} present plots of $\Sigma r^{1/2}$ averaged over 25 orbits for a disc of $\beta = 2$ containing a 33~\earthmass \ protoplanet. The leftmost panel is a 100\% opacity calculation in which the protoplanet exhibits outward migration, whilst the neighbouring panel is for a 1\% opacity calculation that leads to inward migration. Relative to the low opacity case there is a more substantial deficit of material trailing the protoplanet (negative $\phi$) in the horseshoe region of the high opacity case. The rightmost panel of Fig.~\ref{fig:cylindrical} shows a ratio of the two surface density perturbation plots, and makes it clear that there is a greater density contrast between the leading and trailing positions in the horseshoe region of the high opacity outwardly migrating case. This density structure is reminiscent of that found in the two-dimensional models, though much less clearly defined.

\section{Summary}
\label{sec:summary}

We have conducted a series of calculations to explore the impact of disc radial temperature profiles on the rates of protoplanet migration. The surface density profile remains unchanged in each calculation, thus each temperature profile gives rise to a different entropy profile through the disc. In agreement with analytic predictions, we find that steeper temperature profiles, $\beta >1$, (and so entropy gradients) are much more disposed to bring about outwards migration for low mass ($\lesssim 50$~\earthmass) protoplanets. There remains a dependence on the disc's opacity, which determines how the disc and the embedded protoplanet are able to cool, which can change the discs structure and the resulting torques.  In interstellar grain opacity (low thermal diffusivity) discs, low mass protoplanets are found to migrate outwards, whilst in discs with opacities of 1\% this level we find no cases of outward migration. The dependence of migration rates upon the disc's temperature profile in high opacity discs is in agreement with the predictions of recent analytic work by \cite{MasCas2010}.

For the low opacity calculations we find migration rates broadly inline with the rates predicted by \cite{TanTakWar2002}. These conditions can be represented using the formulae of \cite{MasCas2010} with a suitably high thermal diffusivity. Such an approach suggests that a given protoplanet should migrate more rapidly inwards in a disc with a more steeply negative temperature profile. We found this to qualitatively agree with recent work by \cite{DAnLub2010} who find a similar trend for a locally-isothermal disc. However, our efforts to numerically test these predictions were hampered by large numerical uncertainties in the results of our low opacity calculations, leaving us unable to determine the veracity of the analytic predictions.

We also expanded our investigation into the importance of the model used to represent the protoplanet. Using a surface that enables accreted gas to pile up and form an atmosphere results in somewhat different migration rates to those obtained when using a more crude accreting sink particle model. For the lowest-mass protoplanets ($\lesssim 20$~\earthmass) the change in migration rate with protoplanet treatment was found to be small, though for borderline cases between inward and outward migration it could be sufficient to change the direction. Those protoplanets of $\gtrsim 33$~\earthmass \ using the less realistic accreting sink model are found to migrate more rapidly inwards as they more effectively clear the coorbital region of material than equivalent surface models, more rapidly diminishing the positive coorbital torques. In general the effect of changing the protoplanet treatment is small compared to the differences brought about by altering the disc properties.

\section*{Acknowledgments}

We thank the anonymous referee for prompting some further useful investigation. The calculations reported here were performed using the University of Exeter's SGI Altix ICE 8200 supercomputer. Much of the analysis was conducted making use of SPLASH \citep{Pri2007}, a visualisation tool for SPH that is publicly available at http://www.astro.ex.ac.uk/people/dprice/splash. MRB is grateful for the support of a Philip Leverhulme Prize and a EURYI Award which also supported BAA. This work, conducted as part of the award ``The formation of stars and planets: Radiation hydrodynamical and magnetohydrodynamical simulations"  made under the European Heads of Research Councils and European Science Foundation EURYI (European Young Investigator) Awards scheme, was supported by funds from the Participating Organisations of EURYI and the EC Sixth Framework Programme.

\bibliography{paper.bib}

\end{document}